# The Deployment of an Enhanced Model-Driven Architecture for Business Process Management

Richard McClatchey

*Centre for Complex Cooperative Systems, University of the West of England, Bristol, UK*
*Richard.McClatchey@uwe.ac.uk*

**Abstract:** Business systems these days need to be agile to address the needs of a changing world. Business modelling requires business process management to be highly adaptable with the ability to support dynamic workflows, inter-application integration (potentially between businesses) and process reconfiguration. Designing systems with the in-built ability to cater for evolution is also becoming critical to their success. To handle change, systems need the capability to adapt as and when necessary to changes in users' requirements. Allowing systems to be self-describing is one way to facilitate this. Using our implementation of a self-describing system, a so-called description-driven approach, new versions of data structures or processes can be created alongside older versions providing a log of changes to the underlying data schema and enabling the gathering of traceable ("provenance") data. The CRISTAL software, which originated at CERN for handling physics data, uses versions of stored descriptions to define versions of data and workflows which can be evolved over time and thereby to handle evolving system needs. It has been customised for use in business applications as the Agilium-NG product. This paper reports on how the Agilium-NG software has enabled the deployment of an unique business process management solution that can be dynamically evolved to cater for changing user requirement.

**Keywords:** Description-driven systems, business process management, business provenance, system evolution

## 1  INTRODUCTION

With the growth in complexity of global marketplaces organizations are increasingly pressured to re-structure, diversify, consolidate and often re-focus to provide a competitive edge. Furthermore, with the advent of Cloud computing and the need for interconnection and cooperation (and often coexistence) with legacy systems, the demand for flexible and adaptable software has increased. When this is coupled with increasing system complexity and the requirement for systems to evolve over long timescales, the importance of clearly defined, extensible models for rapid systems design becomes crucial. Systems need to be designed to handle on-going change in a dynamic and seamless manner so that they can react and adapt to changes in requirements as they emerge over time.

   BPM [1] tools have been proposed to handle the problem of harmonizing different cooperating business systems. The object of these business orchestration tools is to allow technical teams to collaborate to define and optimize processes by using a single tool to coordinate activities across multiple processes. However, these tools often do not have the ability to enable dynamic design changes to be rapidly reflected into production-ready systems. Few BPM solutions offer the possibility to react to changes in the business environment of the process whilst in operation. Even fewer are able to intervene in executing processes, eliminating or skipping steps along the way or adding new steps to replace or supplement existing execution. What is needed therefore are design approaches that allow systems to reconfigure dynamically, so that designs can be versioned and rolled out into production to sit seamlessly alongside existing production systems without intervention or downtime. Researchers in the Centre for Complex Cooperative Systems (CCCS) at the University of the West of England (UWE) have implemented the CRISTAL system (as described in [2]), based on data and meta-data descriptions that enable systems to dynamically reconfigure and to have system descriptions managed alongside collected provenance data. CRISTAL has been used as the basis of the Agilium-NG product (www.agilium.com [3]) to provide the ability to capture system semantics and have those semantics 'discovered' and used by external systems.

   The next section shows how change management can be facilitated by a model-driven design approach and summarises what we term the *description-driven systems* (DDS) philosophy, which has been developed to address the needs of enterprises that need to respond to changing processes. We then detail the CRISTAL DDS [2], that allows businesses to handle change dynamically whilst enabling all enterprise components to co-operate. A report of its use in a BPM system [4] over time is presented, identifying benefits for modelling and data/process traceability. Conclusions are drawn, and indications given as to future work that can facilitate a new generation of enterprise modelling.

## 2  MODEL-DRIVEN DESIGN AND DESCRIPTION-DRIVEN SYSTEMS

There is a growing body of research that shows that a model-driven approach to systems design provides for both short-term and long-term productivity [5] for developers. Atkinson and Kuhne [6] have shown that meta-modelling is an essential foundation for model-driven development (MDD) and that MDD reduces the sensitivity of systems to change. They advocate that the traditional four-level OMG modelling infrastructure (Data-Model-Metamodel and Meta-metamodel levels [7]) should be based on a two-dimensional approach to meta-modelling involving both 'linguistic meta-modelling' (using linguistic "instances-of" relationships) and ontological meta-modelling (ontological "instances-of" relationships). They propose that a disciplined use of meta-modelling and object-oriented (OO) can lead to heterogeneous, extensible and open systems. Our research [8] has shown that use of meta-modelling creates flexible systems offering reusability, complexity handling, version handling, system evolution and inter-operability. Consequently, meta-modelling is a powerful and useful technique in designing domains and developing dynamic systems.

There are a number of OO design techniques that encourage the design and development of extensible and reusable objects. In particular design patterns are useful for creating reusable OO designs [9]. Design patterns for structural, behavioural and architectural modelling have been well documented elsewhere and have provided software engineers with rules and guidelines that they can use and reuse in software development. Reflective architectures that can dynamically adapt to new user requirements by storing descriptive information which can be interpreted at runtime have led to so-called Adaptive Object Models [10]. These are models that provide meta-information about domains that can be changed on the fly. Such an approach, proposed by Yoder, is similar to that adopted in this paper.

A reflective system uses an open architecture where implicit system aspects are reified to become explicit first-class meta-objects [11, 12]. The advantage of reifying system descriptions as objects is that operations can be carried out on them, like discovering and organizing, composing and editing and storing and retrieving. Since these meta-objects can represent system descriptions, their manipulation can result in changes in the overall system behaviour. As such, reified system descriptions can lead to dynamically evolvable and reusable systems. Meta-objects, as used in the research reported in this paper, are the representations of the system describing how its internal elements can be accessed and manipulated. These representations are causally connected to the internal structures they represent, so changes to these representations immediately affect the underlying system. The ability to dynamically augment, extend and redefine system specifications results in a considerable improvement in flexibility. This leads to dynamically modifiable systems, which can adapt and cope with evolving requirements.

We propose a design approach based on a model-driven philosophy that optimises flexibility and evolution to facilitate adaptive systems modelling. We call our approach *description-driven*. It involves abstracting, at the outset of design, all the crucial enterprise model elements (such as business objects, processes, lifecycles, goals, agents and outputs) and creating high-level descriptions of these elements which are stored in an enterprise model, that can be dynamically modified. A Description-Driven System (DDS) architecture [2, 11] makes use of meta-objects to store domain-specific system descriptions, which control and manage the life cycles of meta-object instances, i.e. domain objects. In a DDS the separation of descriptions from their instances allows them to be specified and managed and to evolve independently and asynchronously. This separation allows the realization of inter-operability, reusability and system evolution as it gives a clear boundary between the application's basic functionalities from its representations and controls.

The DDS architecture is an extension of the four-layer (Data, Model, Meta-Model and Meta-Meta-Model) architecture and it follows the two-dimensional approach of Atkinson & Kuhne [6]. To aid comprehension, we present here how the relational mode, would be represented as a DDS (figure 1). The vertical abstraction of the DDS architecture is based on "instance-of" relationships. The data layer corresponds to the objects that are manipulated in the system. In a DDS, we add a horizontal abstraction encompassing a "describes" relationship for each of the four layers of the model-driven architecture that enables the meta-data for each horizontal layer. The type of meta-data that can be extracted in the data layer concerns the physical aspects of the data such as the volume or localization. The model layer corresponds to concepts for model description that describe data and meta-data in the data layer. In the relational model the meta-data that can be extracted from the model layer are descriptive information concerning the physical structure, e.g. data dictionary or catalogue.

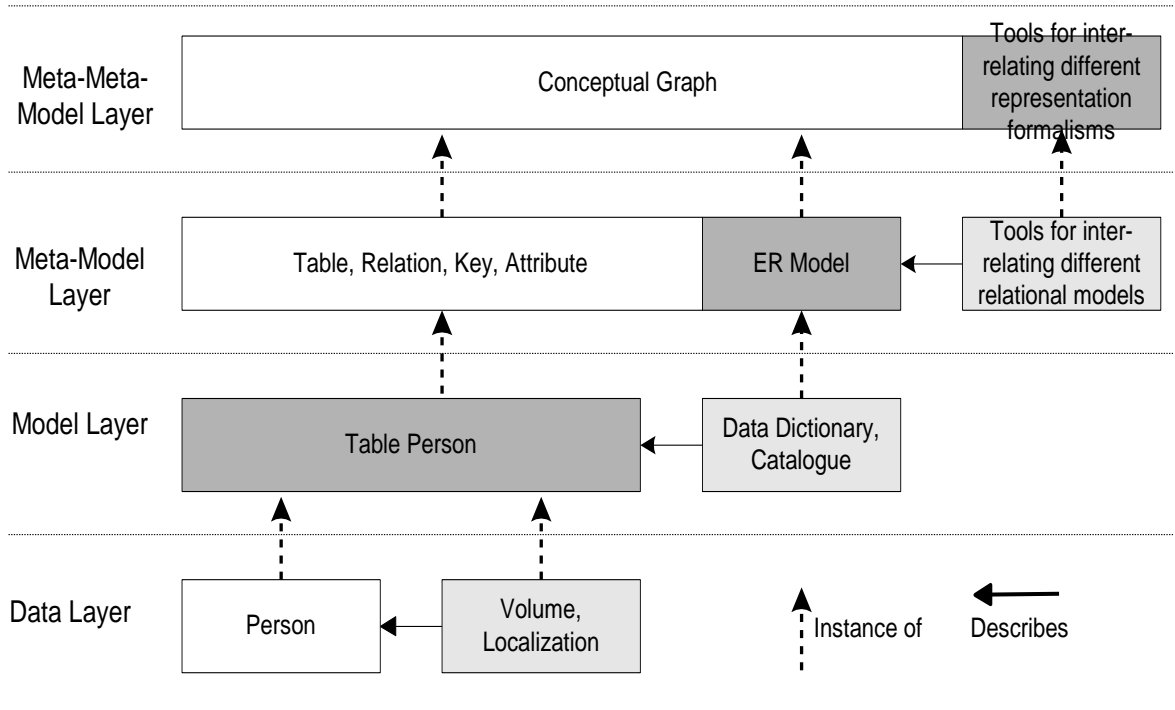

**Figure 1: Relational Model represented as an example of a Description-Driven System**

The meta-model layer defines the modeling formalism that is used in the system. In this example the architecture is intended for relational systems, so the meta-model layer describes the concepts of the relational model, e.g. Table, Relation, Attribute and Key and the model that relates these primitives together. At this level, the type of meta-data that can be extracted concerns tools and methods for inter-operating and relating the different relational models. The meta-meta-model layer is the root modelling level that supports inter-operability and extensibility. In the example of the relational model, the meta-meta-model layer uses conceptual graphs [13] to allow homogeneous representation and manipulation of the lower levels. Conceptual graphs are formalisms where the universe of discourse can be modelled by concepts and conceptual relations and are fundamental to the relational model.

Our DDS architecture is akin to the architecture described in [14] but is based on an object model. Horizontal layers are based on model abstraction, and meta-data are used as descriptive information to describe the concepts for managing data at each layer. The two architectures are orthogonal in the modelling paradigms used - relational model for [14] and object model for DDS. At the meta-meta model layer [14] uses conceptual graphs and the DDS architecture uses the MOF [7]. In the meta-model layer, [14] uses a relational model and the DDS architecture uses the UML [7]. As a practical example of our approach the next section describes the DDS architecture developed in the context of research carried out in the data management CRISTAL project at CERN and later by the Agilium-NG system for BPM. A DDS, as defined in detail in [2], makes use of *meta-objects* to store system descriptions, which manage the lifecycles of instances of those meta-objects. The meta-object instances are the managed domain objects, such as the activities in a BPM system, but could also be data objects, documents, role definitions or any business-critical item that requires management over time.

## 3  CRISTAL: A DDS FOR MANAGING CHANGE

We originated the DDS approach for use at CERN where scientists build and operate complex experiments whose construction processes are very data-intensive, highly distributed and require computer-based systems to orchestrate and trace complex workflows. In constructing experiments like the Compact Muon Solenoid (CMS, [15]), scientists required data management systems that could cope with complexity, with system evolution, with ease of management over time and with system scalability. These are the same problems that are currently being faced by industry in delivering the next generation of BPM systems. CMS is a general-purpose high-energy physics experiment that has been constructed from around a million parts and assembled over a decade by specialized centres distributed worldwide. In its construction process its production models rapidly evolved but data gathering was required to be 'round-the-clock' with minimal system down time. Experiment components defined by model versions needed to be handled over time and to coexist with other components from different model versions. Separating details of model types from the details of parts allowed the model type versions to be specified and managed independently, asynchronously and explicitly from single parts, even while the research processes themselves evolved over time.

A research project, entitled CRISTAL was initiated to facilitate the management of the data collected at each stage of development of CMS. CRISTAL is, in essence, a distributed product data and workflow management system which makes use of an object database for its repository, a multi-layered (DDS) architecture for its component abstraction and dynamic object modelling for the design of the objects and components of the system [8]. The DDS approach has been followed to handle the complex data-intensive systems and to provide the flexibility to adapt to changing research scenarios. This paper does not dwell on the detailed technical aspects of the CRISTAL DDS model; a full description can be found in [2], but for clarity figure 2 reproduces its multi-level architecture.

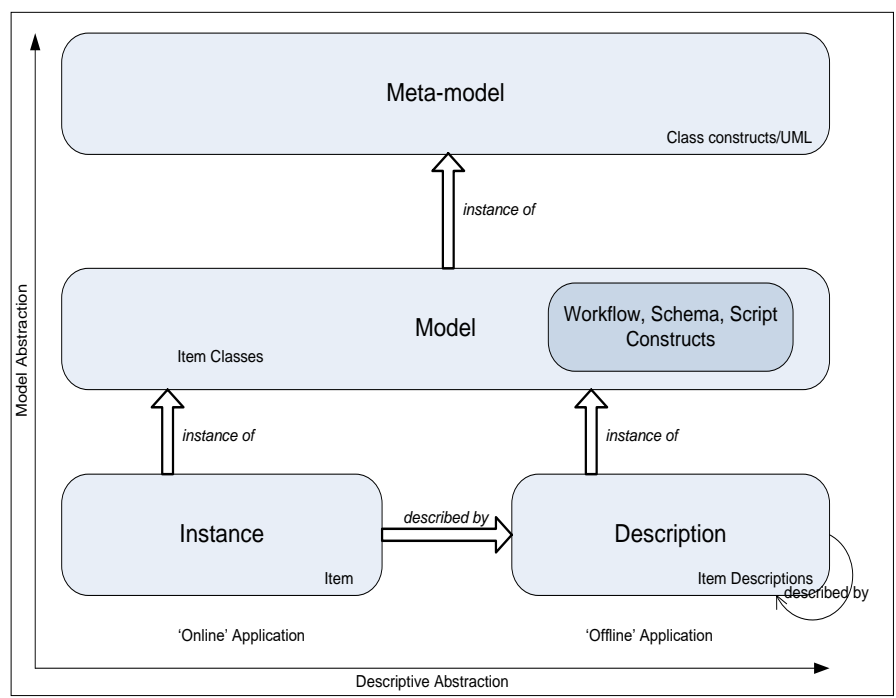

**Figure 2: Model versus Description in CRISTAL (from [2])**

Within CRISTAL each defined system element (or Item) is stored and versioned. This allows users of the system to view older versions of their Items and either extend a version of an Item or return to an earlier version of an Item. In CRISTAL a model is designed as a set of Items, each having a lifecycle defined by workflows, themselves composed of activities. The activities comprising an Item's lifecycle have similarities with object-oriented methods, since they define a single action performed on that Item. Each activity set must always form a valid graph of activities from the creation of the Item to its completion. This clarity of design through implementation constraints is a return to the philosophy advocated in early OO languages such as Smalltalk [16], which steered towards a core design with the system logic partitioned in a manageable way.

*Item Description* Items hold the templates for new Items and also dictate their type (see Figure 2). These "Item Descriptions" are themselves declared as Items (and thus the two can be treated in the same manner with common code), holding the description data as XML outcomes managed through workflow activities. Workflow and Property descriptions are stored as XML serialized objects. Collection Descriptions are themselves Collections, pointing to other Item Descriptions. Outcome Descriptions contain XML Schema documents which are used to validate submitted outcomes and aid in data collection, for instance to generate data entry forms in a stock GUI for the end users. Also included in the descriptions are Scripts, code invoked by workflows either during a change of Activity state to enact consequences of the execution such as updating a Property or changing a Collection, or to assess conditional splits in the Workflow. During the nine years of near-continuous operation of CRISTAL at CERN, the descriptions went from beta to production then through years of relatively few alterations of the domain logic which necessitated very limited change in the actual server software, illustrating the flexibility of the CRISTAL approach (see Table 1). These alterations were minor and included updates to descriptions of processes and data sources which were handled by version management capability of CRISTAL. The server software only needed to be upgraded seven times, and of those seven, only one was a required update that needed to be made available to all users and servers.

| Table 1 - Statistics of CRISTAL operation at CERN | |
|---|---|
| Total number of centres (servers) | 9 (6 at CERN, 1 in Taiwan, 2 in Greece) |
| Runtime | 9 years |
| Total data size (at CERN) | 0.4 TB |
| Total number of Items in one ECAL | 450,000 |
| Minor version upgrades (required client update) | 1 |
| Total number of kernel builds | 22 |
| Kernel builds requiring server software upgrade | 7 |

In CRISTAL improvements can be quickly performed online, often by modifying the workflow of one test item, which then serves as a template for the type definitions. Items subject to the improvements can co-exist with items generated earlier and prior to the improvement being made and both are accessed in a consistent, reusable and seamless manner. All this can be done without recompiling a single line of code or restarting the application server, providing significant savings in time and enables the users to work in an iterative and reactive manner that suits their research. This shows the flexibility of using a DDS approach.

In practice, following a DDS approach allowed the CRISTAL to be open in design and flexible in nature. Its design simplicity was not compromised by being viewed from one or several application-led standpoints. Rather we concentrated on the traceability of the essential enterprise objects over the lifetime of the system as the primary goal and left the application-specific views to be defined as and when they became required. These enterprise objects each had a creation/modification/deletion lifecycle and the CRISTAL model keeps track of status changes to its Items over those lifecycles. This allows it to orchestrate the execution of workflows on Items by agents, to log all events, outcomes and decision points and thereby capture all associated provenance information associated with any domain system under study. The ability of description-driven systems to both cope with change and to provide traceability of such changes (i.e. the 'provenance' of the change [17]) we see as one of the main contributions of the CRISTAL approach to building flexible and maintainable systems; we believe this makes a significant contribution to how enterprise systems can be implemented in the future. Great benefits in terms of maintainability and flexibility resulted from being able to treat many different system objects in a single standardised manner. Savings over the lifetime of the ECAL project at CERN are estimated at several man years of effort.

## 4   DDS AND THE ROLE OF PROVENANCE FOR BUSINESS APPLICATIONS

The in-built ability for a DDS to capture provenance data while system lifecycle activities are being orchestrated provides the functionality required to collect the provenance associated with each Item in the DDS. Provenance information has, in the past decade, been used in digital libraries where its data refers to the documentation of processes in a digital object's life cycle [18]. As explained by Ram and Liu [20], its use has been growing in the computer science domain as provenance has become important in the management of newly generated datasets. Indeed [19] demonstrates that in the scientific world, there is more and more collaboration which leads to a proliferation of shared and reusable data. Provenance data collection and curation has been employed in the life sciences as well, for example in neuroscience [20] or bioinformatics [21]. Moreau and Groth [22] detail five distinct use cases of provenance. By recording the different events that happen to any digital object, provenance enables a high characterization of that object. It gives a clear identity to an object by answering seven W's (What, When Who, Ho(W), Where, Which and Why) [21]. The 'What' specifies the events that happen to the object and the other W's allow specifying the details of the event [22]:

- WHEN: the time at which event occurs
- WHO: the agent responsible for the event
- HOW: the mechanism or actions that caused the event to occur
- WHERE: the place where the event occurs
- WHICH: the instrument that makes event happen
- WHY: the reason the event occurs

In practice the gathering of provenance information enables the construction of a graph describing the relationships among all the DDS elements (its sources, processing steps, contextual information and dependencies) [23]. The Open Provenance Model (OPM) [18] was a first core representation of a standardised provenance model. It has been replaced by the PROV Data Model (PROV-DM), which is defined as "a generic data model for provenance that allows domain and application specific representations of provenance to be translated into such a data model and interchanged between systems" [24]. PROV has now become a W3C (World Wide Web Consortium) standard. The collection of provenance information for business has been proposed by Houy et al [25]. They coin the term *business process provenance* as the "systematic collection of the information needed to reconstruct what has actually happened" in the business process. We note that business provenance can equally well refer to the business data itself as to the business processes. Provenance brings the following design benefits for business applications:

1. System knowledge: thanks to provenance, it is possible to interpret and understand a system-generated result. As explained by Simmhan, Plale and Gannon [26], by analysing the sequence of steps that led to a result, it is possible to gain insights into the chain of reasoning used in its production. For example, in each run of a scientific experiment (with different configurations parameters) there will be specific results. Provenance can help to make the link between results and the corresponding configurations parameters. It therefore provides the context and thus the possible explanation of the result. It adds meaning to the result and can thereby turn the result data into information/knowledge.

2. Reproducibility: provenance enables reproducibility of results. As it captures the characterization of an object or a set of generated results, it is possible to re-obtain exactly the same object/results by following its chain of reasoning and activities in a step-wise fashion.

3. Reliability: according to Cheney et al. [27], one important motivation for using provenance is that it enables assessing authenticity and integrity because provenance provides knowledge of its origin. In terms of quality this could be very useful since it enables tracking back sources of errors. Provenance can be used for the purpose of creating an audit trail to evaluate if any errors were made in processing.

Cubera et al. [28] introduced the term Business Provenance, which uses provenance for the business environment. It is facilitated by a generic data model and middleware infrastructure which can provide traceability of end-to-end operations. Such a technology allows the automatic discovery of what has happened during any business process execution. In their study, the authors demonstrated the use of provenance in the BPM domain with a main argument based around compliance. They explain that business processes rely on human activities and that business operations often differ from their original design, which can lead to business integrity lapses and compliance failures. To avoid supplementary work and costs due to those failures, they propose a compliance solution which tracks business operations. The principle of the Business Provenance technology is to extract the relevant information to address a specific compliance goal. Provenance in this case enables the tracking of end-to-end operations for assuring compliance. Other studies have used this Business Provenance technology to go further in compliance monitoring with the work of Doganata et al. [29] and to enable advanced case management with the work of Martens et al. [30].

A recent investigation undertaken by Uronkarn and Senivongse [31], provides arguments for using the traceability concept in business processes. They propose a methodology which uses traceability to manage business process change. In this study, the traceability is used for improving software development. Indeed, it provides a technology which tracks changes which happen by comparing the current business process model with the newly designed model. This enables users to identify exactly where a development or a modification of the software is required, to implement the changes.

Agilium-NG, a BPM system based on CRISTAL, has been produced by the M1i company based in Annecy, France, to support BPM [32]. As a result of it being based on CRISTAL, Agilium-NG is the only existing BPM software on the market that is able to modify on-going processes, in a graphical form and in real time. Without interrupting the business processes, users can make modifications directly and graphically on any process parameter (flow, roles, GUIs etc). This innovative and unique function guarantees company flexibility by allowing it to react rapidly to external events which have not been modelled without having to review or complicate the model itself. Agilium-NG's deployment is outlined in the next section where a qualitative review of its benefits is considered.

## 5 THE USE OF AGILIUM-NG IN PRACTICE FOR BPM

The CRISTAL system has been adopted and is being commercially exploited by the M1i company for business process management (BPM) and the integration and co-operation of multiple business processes in business-to-business applications. The product, itself called Agilium, addresses the harmonisation of business processes by the use of a CRISTAL database so that multiple, potentially heterogeneous, processes can be integrated with each other and have their workflows tracked in the database. The major area that Agilium-NG uses CRISTAL for is business process capture and recording of their BPM executions. Within Agilium-NG, the model is identical to the internal model of CRISTAL where events are generated and stored. Agilium-NG also integrates the management of data coming from different sources and unites BPM with workflow management, business activity management (BAM) and Enterprise Application Integration (EAI) through the capture and management of their designs in the CRISTAL database [33]. Agilium-NG uses the description-driven nature of the CRISTAL-Agilium-NG models to act dynamically on process instances already running and can thus intervene in the actual process instances during execution. These processes can be dynamically (re-)configured based on the context of execution without compiling, stopping or starting the process and the user can make modifications directly and graphically of any process parameter. The Agilium-NG system aims to provide the level of flexibility for organisations to be agile in responding to the ongoing changes required by cyber-enterprises. Its architecture is shown diagrammatically in figure 3.

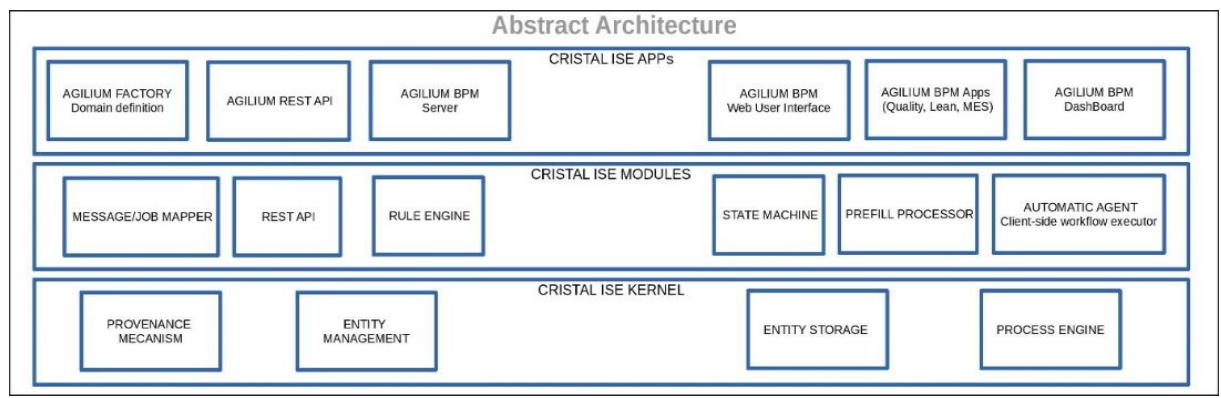

**Figure 3. Agilium-NG's use of the CRISTAL-ISE Kernel**

Using the facilities for description and dynamic modification in CRISTAL, Agilium-NG can provide a modifiable and reconfigurable workflow system. The workflow description and enactment elements of CRISTAL are correlated and each instance stores any modifications that have been carried on it. With an efficient model of verification at both levels (i.e. description and enactment) it is possible to validate if the migration from one description to another one within an instance is possible and to detect any modifications and therefore to apply the migration. Innovating technologies used in the kernel of CRISTAL provides significant advantages when compared to alternative EAI solutions:

- Flexibility. Architecture independence allows CRISTAL to adapt to new domains and/or new enterprises without any specific development. This is an essential factor in helping to reduce maintenance costs and to minimise conversation costs, thus providing high levels of flexibility.
- Platform independence. Use of JAVA/XML/LDAP technologies allows CRISTAL to work on any preinstalled operating system (Linux, Windows, UNIX, Mac OS) and on any machine.

- Database independence. XML storage with LDAP referencing makes CRISTAL autonomous and independent of any underlying database and of any application server.
- Simplified integration of applications. XML presents a solution that supports multiple heterogeneous environments.
- Fully distributed. This functionality provides web usability through the Internet or an Intranet. It also makes data accessible from multiple databases via a unique interface.

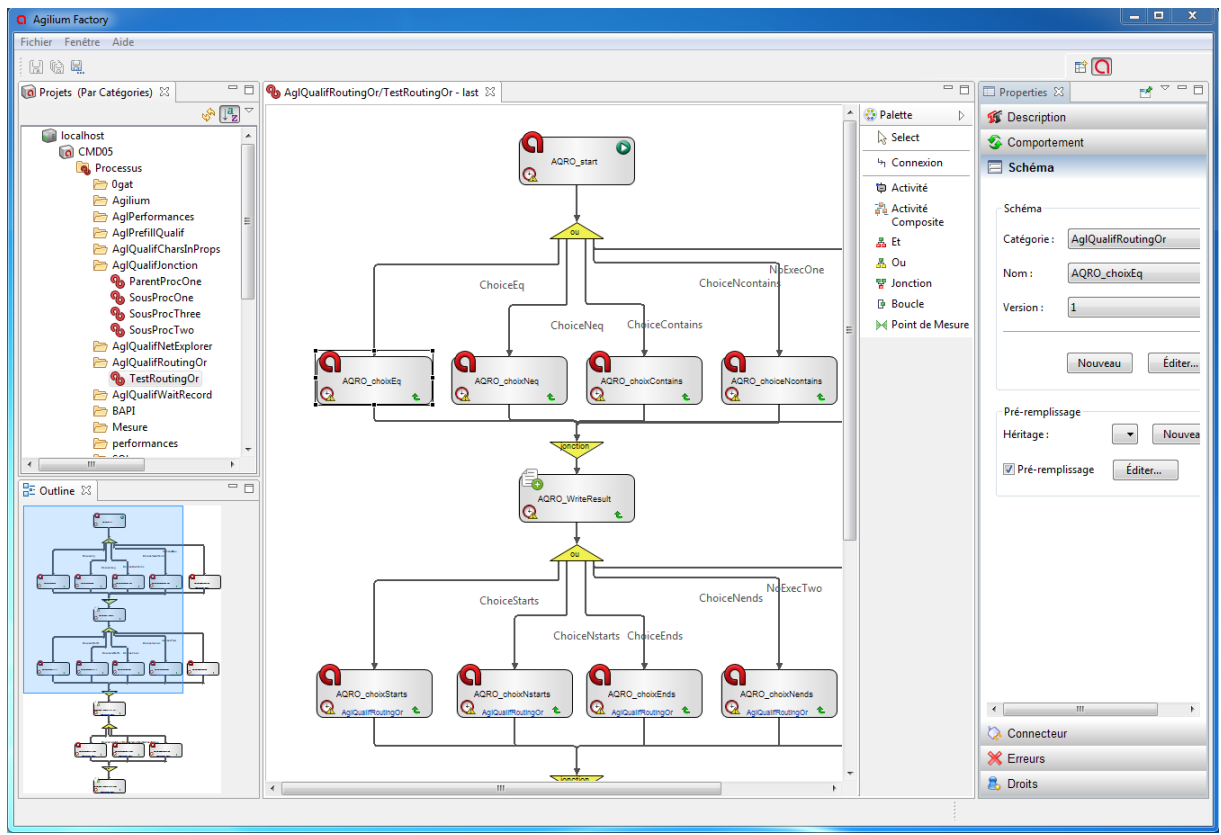

**Figure 4: The Agilium-NG Factory Application.**

The Agilium-NG Server is based on CRISTAL, but with several domain extensions and support for additional protocols added. The user interface (UI) components are the Agilium-NG Web component, the Agilium-NG Supervisor GUI and the Agilium-NG Factory. The Agilium-NG Web is a web application based on J2EE and running within Tomcat as the container. This is where users can browse the currently active jobs and different instances of business processes. The list of jobs available to a user is constrained by their individual roles (for example, administrator). The web UI also allows users to complete manual activities. The supervisor GUI is derived from the original Java Swing CRISTAL GUI and is used by administrators of the system to be able to design and debug workflows and for general system management. The key component in Agilium-NG is known as the Factory. The factory is a full Eclipse-based application which allows M1i's users to create and manage their own CRISTAL based workflows. A screenshot of the Agilium-NG Factory is shown in figure 4.

The Agilium-NG BPM system provides many benefits: firstly, it allows the modification of an ongoing process in a graphical form and in real time. Without interrupting the business processes, the user can make modifications directly and graphically to any process parameter (workflow, data definition, roles). Secondly, on the basis of process indicators and/or piloting rules and external events, the system can modify the planned execution of a process without previous modelling. This function guarantees company flexibility by allowing it to react rapidly to external events which have not been modelled, without having to review or complicate the model itself. The open Source version of CRISTAL (named CRISTAL-ISE after the EU project in which it was developed) is available under the LGPL3 licensing scheme and via GitHub. In Agilium-NG the CRISTAL-ISE kernel can capture information about the application area in which a specific instance is being used, such as the domains of finance, production control, retail management, etc. This can in principle allow usage patterns to be described and captured, roles and agents to be defined on a per-application basis and rules and outcomes specific to user domains to be managed. In turn this enables multiple instances of Agilium-NG to discover the semantics needed for them to inter-operate and to exchange data. This should have a profound effect in terms of business-to-business operation and the ease of maintainability of systems involving multiple instances of Agilium-NG. Further details of the Agilium-NG product can be found at [3].

## 6  CONCLUSIONS AND FUTURE DIRECTIONS

Much progress has been made in the last decade in identifying approaches that can inform the process of model-driven software evolution [34]. There have been many model-driven systems that have emerged that go some way to providing support for evolving systems; examples of these include ERP/BPM systems such as ADempiere [35], Whitestein LSPS [36] and Compiere [37]. All of these systems allow, to some extent, users to reformulate workflows and data structures on-the-fly, and provide varying levels to control the management of changes, but they all implement their meta-models entirely separately from their instances. In the approach proposed in this paper both the descriptions and the instances of those descriptions are implemented as objects and are maintained using the same software. Even though workflow descriptions and instance implementations differ in content, the way they are stored and relate to each other is fundamentally the same in CRISTAL.

The CRISTAL DDS architecture was shown to have two abstractions. The vertical abstraction is based on the "is an instance of" relationship from the OMG meta-modelling standard and has (at least) three layers - instance layer, model layer and meta-model layer. This paper has proposed an orthogonal horizontal abstraction mechanism that complements this approach. The horizontal abstraction is based on the meta-level architecture approach, encompasses the "is described by" relationship and has two layers - meta-level and base level. This description-driven philosophy facilitated the design and implementation of the CRISTAL project with mechanisms for handling and managing reuse in its evolving system requirements and served as the basis of the Agilium-NG description-driven BPM workflow system.

The studies reported in this paper have shown that describing a system explicitly and openly from the outset of the project, using the DDS approach, enables the developer to change aspects of it responsively as users' requirements evolve. This allows seamless transition from one version to the next with (virtually) uninterrupted system operation and facilitates traceability via provenance data collection throughout the system lifecycle. Following the principles of OO design the DDS approach encourages reuse of code and scripts/methods. Indeed, the description-driven design approach takes this one step further and provides reuse of meta-data, design patterns [11] and maintenance of items and activities (and their descriptions). Practically this results in a higher level of control over design evolution and simpler implementation of system improvements in addition to simpler maintenance cycles. Many system elements have consequently gained in conceptual simplicity and ease of management thanks to loose typing and the adoption of a unified approach to their manipulation. One consequence of providing such a unified design and simplicity of management is that the CRISTAL software can be used for a wide spectrum of domains. The contributions of this work complement the ongoing research on Adaptive Object Models (AOM) expounded in [10] and [38], where a system with an AOM (also called a Dynamic Object Model) is stated to have an explicit object model that is stored in the database and interpreted at runtime.

Future work is being pursued to model domain semantics e.g. the specifics of a particular application domain e.g. retail, finance, and aerospace. This will essentially transform CRISTAL into a self-describing model execution engine, making it possible to build applications directly on top of the design, without code generation. The design will be the framework for all of the application logic – without the risks of misalignment and subsequent loss that code generation can bring – and for CRISTAL to be configured as needed to support the application logic whatever it may be. What this means is that the CRISTAL kernel will be able to capture information about the application area in which a particular instance is being used. This will allow usage patterns to be described and captured, roles and agents to be defined on a per-application basis, and rules and outcomes specific to particular user domains to be managed. It is also planned to investigate how the semantics of CRISTAL items and agents could be captured in terms of ontologies and thus mapped onto or merged with existing ontologies for the benefit of new domain models. This will enable multiple instances of CRISTAL to discover the semantics required to inter-operate and to exchange data. This should have a profound effect in terms of business-to-business operation and the ease of configuration and maintainability of systems involving multiple instances of CRISTAL. Thus, we can enable business to business (B2B) collaboration by, for example, allowing one company using CRISTAL/Agilium-NG to discover information from another company also using CRISTAL/Agilium-NG and thereby to facilitate cooperation.

In future work a provenance analysis mechanism will be built on top of the data that has been captured in CRISTAL. It will employ software to learn from the data that has been produced, classify and reason from the information accumulated and present it to the system in an intuitive way. This information will be delivered to users while they work on new business process workflows and will be an important source for future decision-making. One essential future element is the provenance interoperability aspect. Currently, we are working on exporting the provenance enabled objects to the emerging PROV [22] interoperability standard. This will allow businesses to use their provenance data in other PROV-compliant systems.

## ACKNOWLEDGEMENTS

The author wishes to highlight the support of his home institute and acknowledges the support of the European Union for the CRISTAL-ISE project under the 2011-2012 Marie Curie Industry and Academic Pathways Partnership (IAPP) scheme contract number: 285884.

## REFERENCES

[1] J. Jest on & J. Neil's (2006), *Business Process Management. Practical Guidelines to Successful Implementation*. Butterworth-Heinemann publishers, 2006.

[2] A. Branson et al. (2014), "CRISTAL : A Practical Study in Designing Systems to Cope with Change". *Information Systems 42,* pp 139-152 http://dx.doi.org/10.1016/ j.is.2013.12.009. Elsevier publishers.

[3] Agilium-NG product. See http://www.agilium.com Last accessed March 2018.

[4] M. Weske "*Business Process Management. Concepts, Languages, Architectures*". Springer publishers, 2007.

[5] A. van Deursen et al. (2007), "Model-Driven Software Evolution: A Research Agenda*". Proceedings of the 11th European Conference on Software Maintenance and Reengineering*. Workshop on Model-Driven Software Evolution - MoDSE2007. Amsterdam, the Netherlands.

[6] C. Atkinson and T. Kuhne (2003) "Model-Driven Development: A Meta-modelling Foundation". *IEEE Spectrum* Vol 20 No. 5 pp 36-41. September 2003. IEEE Press

[7] The Object Management Group (OMG), URL http://www.omg.org. Last accessed March 2018.

[8] F. Estrella et al. (2001), "Meta-Data Objects as the Basis for System Evolution". *Lecture Notes in Computer Science Volume 2118*, p. 390-399 ISBN 3-540-42298-6 Springer-Verlag.

[9] E. Gamma, R. Helm, R. Johnson and J. Vlissides (1995), *Design Patterns: Elements of Reusable Object-Oriented Software*, Addison-Wesley.

[10] J. Yoder and R. Johnson (2002), "The Adaptive Object Model Architectural Style". *Proceedings of the Working IEEE/IFIP Conference on Software Architecture* 2002 (WICSA3 '02)

[11] F. Estrella et al. (2003), "Pattern Reification as the Basis for Description-Driven Systems". *Journal of Software and System Modelling* Volume 2 Number 2, pp 108-119 Springer-Verlag, 2003.

[12] G. Kiczales (1993), "Metaobject Protocols: Why We Want Them and What Else Can They Do?", Chapter in *Object-Oriented Programming: The CLOS Perspective*, pp 101-118, MIT Press.

[13] J. Sowa, (1984) *Conceptual Structures: Information Processing in Mind and Machine*, Addison-Wesley.

[14] B. Kerherve and O. Gerbe (1997), "Models for Metadata or Metamodels for Data", *Proceedings of the Second IEEE Metadata Conference*, Maryland, USA, September.

[15] S. Chatrchyan et al. (2008), "The CMS Experiment at the CERN LHC". The CMS Collaboration, *Journal of Instrumentation* Vol 3 361 pages IoP Publishers.

[16] A. Goldberg, A et al, *Smalltalk-80: the Language and its Implementation*. Addison-Wesley Longman Publishing Co. 1983.

[17] Y. Simmhan et al., (2005) "A Survey of Data Provenance in e-Science". In *ACM SIGMOD Record, Vol 34*, P. 31-36.

[18] L. Moreau, B. Cliffort, et al (2011) "The Open Provenance Model Core Specification (v1. 1)". *Future Generation Computer Systems*, 27(6), pp. 743-756

[19] S. Ram and J. Liu, (2012) "A Semantic Foundation for Provenance Management". *Journal of Data Semantics* 1(1), pp. 11-17.

[20] A. Anjum et al. (2011) "Provenance management for neuroimaging workflows in neugrid". *Proceedings of the IEEE P2P, Parallel, Grid, Cloud and Internet Computin International Conference* (3PGCIC), pp. 67-74.

[21] C. Goble (2002) "Position statement: Musings on provenance, workflow and (semantic web) annotations for bioinformatics". *Workshop on Data Derivation and Provenance*, Chicago, October 2002.

[22] L. Moreau, L. and P. Groth, (2013) *Provenance: An Introduction to Prov*. San Rafael, California: Morgan & Claypool Publishers.

[23] L. Carata, S. Akoush, N. Balakrishnan et al. (2014). "A Primer on Provenance". *Communications of the ACM*, 57(5), pp. 52-60.

[24] The World Wide Web Consortium W3C (2013. "PROV-DM: The PROV Data Mode"l [online] Available from: https://www.w3.org/TR/prov-dm/

[25] C. Houy, P. Fettke et al., (2010) "BPM-in-the-Large – Towards a Higher Level of Abstraction in Business Process Management". *E-Government and E-Services / Global Information Systems Processes GISP2010 Workshop at the World Computer Congress WCC-2010*. Brisbane, Australia.

[26] Y. Simmhan, B. Plale and D. Gannon (2005) "A Survey of Data Provenance Techniques". Computer Science Department, Indiana University, Bloomington IN, 47405

[27] J. Cheney, S.Chong, N. Foster et al.(2009) "Provenance: a future history". *Proceedings of the 24th ACM SIGPLAN conference companion on Object oriented programming systems languages and applications*, pp. 957-964.

[28] F. Curbera, F., Doganata, et al. (2008). "Business Provenance–a Technology to Increase Traceability of End-to-End Operations". *Proceedings of the OTM Confederated International Conferences" On the Move to Meaningful Internet Systems"*, pp. 100-119.

[29] Y. Doganata, K Grueneberg et al. (2011) "Authoring and Deploying Business Policies Dynamically for Compliance Monitoring". In *Policies for Distributed Systems and Networks (POLICY),* pp. 161-164.

[30] A. Martens, A. Slominski, et al. (2012) "Advanced Case Management Enabled by Business Provenance". *Proceedings of the International Conference on Web* Services (ICWS), pp. 639-641

[31] W.Uronkarn and T. Senivongse (2014) "Change Pattern-driven Traceability of Business Processes". *Proceedings of the International MultiConference of Engineers and Computer Scientists*.

[32] J. Shamdasani et al. (2014) "CRISTAL-ISE: Provenance applied in industry*". Proceedings of the International Conference on Enterprise Information Systems* (ICEIS) Vol 3, pp 453-458

[33] D. Georgakopoulos et al. (1995), "An Overview of Workflow Management", *Journal of Distributed and Parallel Database Systems* 3 (2), pp119-153.

[34] A. van Deursen et al. (2007), "Model-Driven Software Evolution: A Research Agenda*". Proceedings of the 11th European Conference on Software Maintenance and Reengineering*. Workshop on Model-Driven Software Evolution - MoDSE2007. Amsterdam, the Netherlands.

[35] ADempiere ERP. See http:// http://www.adempiere.com. Last accessed March 2018

[36] Whitestein Living Systems Process Suite. See http:// http://www.whitestein.com/products/process-suite/. Last accessed March 2018

[37] Compiere ERP Software. See http://www.compiere.com. Last accessed March 2018

[38] P. Matsumoto et al. (2011) "AOM Metadata Extension Points*" Proceedings of the 18th Conference on Pattern Languages of Programs* (PLoP), Portland, Oregon, USA.